\documentclass[prl,aps,psfig,preprintnumbers,amsmath,amssymb,showpacs]{revtex4}
\usepackage{color,graphicx}\usepackage{epsfig}

\begin{document}

\title{Ensemble in-equivalence in supernova matter within a simple model}

\author{F. Gulminelli$^{1,2}$ }
\author{Ad. R. Raduta$^{3}$}

\affiliation{$^{1}$~CNRS, UMR6534, LPC ,F-14050 Caen c\'edex, France\\
$^{2}$~ENSICAEN, UMR6534, LPC ,F-14050 Caen c\'edex, France\\
$^{3}$~NIPNE, Bucharest-Magurele, POB-MG6, Romania
}

\begin{abstract}
A simple, exactly solvable statistical model is presented for the description
of baryonic matter in the thermodynamic conditions associated to the evolution
of core-collapsing supernova. It is shown that the model presents a first
order phase transition in the grandcanonical ensemble which is not observed in
the canonical ensemble. 
Similar to other model systems studied in condensed matter physics, this
ensemble in-equivalence is accompanied by negative susceptibility and
discontinuities in the intensive observables conjugated to the order parameter.
This peculiar behavior originates from the fact that baryonic matter is
subject to attractive short range strong forces as well as repulsive long
range electromagnetic interactions, partially screened by a background of
electrons. As such, it is expected in any theoretical treatment of nuclear
matter in the stellar environement. Consequences for the phenomenology of 
supernova dynamics are drawn. 

\end{abstract}

\today

\maketitle

\section{I. Introduction}
Standard thermodynamics is based on the assumption that the physical 
properties of a system at equilibrium do not depend on the statistical
ensemble which is used to describe it. Under this condition thermodynamics 
is unique and the different thermodynamic potentials are related via simple
linear Legendre transforms.
Is is however well known that the  equivalence between the different
statistical ensembles can only be proved \cite{vanhove} at the thermodynamic
limit and under the hypothesis of short range interactions, while non-standard 
thermostatistic tools have been developed during the years to deal with 
non-extensive and long-range interacting systems \cite{tsallis,dauxois}.
 
The issue of ensemble in-equivalence, namely the possible dependence 
of the observed physics on the externally applied constraints, 
has been typically associated to 
phase transitions, more precisely to phase separation quenching due to the 
external constraint. A well-known example in the literature concerns the 
possible occurrence of negative heat capacity in finite systems, which has
been widely studied theoretically \cite{gross} and has also given rise to 
different experimental applications in nuclear and cluster physics
\cite{cneg_exp}. In this specific example the phase separation is  quenched 
by the microcanonical conservation constraints, leading to the thermodynamic 
anomaly of a non-monotonous equation of state.

Concerning macroscopic systems, different model applications have shown
fingerprints of ensemble in-equivalence 
\cite{dauxois,ruffo,bouchet,orlandini,mukamel,pre} 
but phenomenological applications are scarce.
In this paper we show that the dense matter which is produced in the explosion
of core-collapse supernova and in neutron stars is an example of a physical 
system which displays this in-equivalence.

We will limit our discussion to finite temperature $T\approx10^{10}K$ 
and nuclear sub-saturation $10^{10}<\rho<10^{14}$ g cm$^{-3}$ densities, 
thermodynamic conditions which are known to be largely explored in the
dynamics of supernova matter and in the cooling phase of proto-neutron stars 
\cite{kitaura,marek}. 
The baryonic component of this stellar matter is given by a statistical
equilibrium of neutrons and protons, the electric charge of the latter being 
screened by an homogeneous electron background. 

If the electromagnetic interactions are ignored, this gives the standard model 
of nuclear matter, which is known to exhibit first and second order phase
transitions with baryonic density as an order parameter, 
meaning that the transition concerns a separation between a dense (ordered) 
and a diluted (disordered) phase \cite{NM}. It is however known since decades 
to the astrophysical community that the situation is drastically different 
in stellar matter, where microscopic dishomogeneities are predicted at almost 
all values of temperature, density and proton fraction and thermodynamical 
quantities continuously change at the phase transition 
\cite{prakash_science,haensel_book}. This specific situation of stellar matter 
respect to ordinary nuclear matter has been shown to be due to Coulomb
frustration which quenches the first order phase transition
\cite{ising_star}. 
However, the thermodynamic consequences of this specific thermodynamics with 
long range interactions have never been addressed to our knowledge.

In this work we will show that these dishomogeneities imply ensemble
in-equivalence, making neutron star matter the first astrophysical example to our knowledge
of ensemble in-equivalence at the thermodynamic limit. 
Additionally, we will show that a consistent treatment of this specific
thermodynamics can have sizeable effects in the equations of state which are
currently used to describe the supernova phenomenology.  

In a recent paper \cite{starmatter_mmm} we have proposed a phenomenological
hybrid model for supernova matter which was numerically solved by Monte-Carlo
simulations. 
Since convergence is always an issue in Monte-Carlo calculations, 
we propose in this paper an analytic version of the same model. 
In order to have analytical results, we will limit to the simplest version of the model where a schematic nuclear energy functional is used. It is clear that more sophisticated energy functionals will have to be implemented in order to have quantitative predictions for the supernova simulations. 

However, the qualitative conclusions of this paper will only depend on the sign (attractive or repulsive) of the interactions and on the 
size dependence (in terms of volume and surface) of the nuclear binding energy. As such, they will not depend on the details of the model.
 
\section{II. The model}

Stellar matter at temperature lower than the typical nuclear binding energy $E_b\approx 8$ MeV/nucleon $\approx 10^{11}$ K, and density $\rho$ lower than 
the saturation density of nuclear matter $\rho_0 \approx 0.16$ 
fm$^{-3} \approx 1.6 \cdot 10^{14}$ g$\cdot$cm$^{-3}$, 
can be viewed as a statistical mixture of free protons and neutrons
with loosely interacting 
nuclear clusters at internal density $\rho_0$, immersed in an homogeneous electron background density $\rho_e$ 
which neutralizes the total positive charge density $\rho_p$ over macroscopic 
length scales, $\rho_e=\rho_p$. Nucleons bound in clusters can be described by a 
phenomenological free energy functional depending on the cluster size $a=n+z$ 
and chemical composition $i=n-z$ as well as on the temperature of the medium:
\begin{equation}
f^\beta_{a,i}=e_{a,i}+\langle e^*_{a,i} \rangle_\beta - Ts^\beta_{a,i} 
\end{equation}
For nuclear (fermionic) clusters both the average cluster excitation energy 
$\langle e^*_{a,i} \rangle_\beta$ and entropy $s^\beta_{a,i}$
can be evaluated in the low temperature Fermi gas approximation
\begin{eqnarray}
\langle e^*_{a,i} \rangle_{\beta}&=& c_0 a T^2; \label{estar_fisher} \\
s^\beta_{a,i}&=& \left ( 2c_0 T +c_S a^{-1/3} h(T) \right ) a ,
\label{entropy_fisher}
\end{eqnarray}
The surface term $c_S a^{2/3}h(T)$ in Eq. (\ref{entropy_fisher}) effectively accounts for the 
entropy increase at finite temperature due to surface excitations,
producing a vanishing surface free energy at a given temperature, 
corresponding to the critical point $\beta_C=T_C^{-1}$ of nuclear matter.
The cluster energy $e_{a,i}$ is modified with respect to the energy of the 
nucleus in the vacuum $e^0_{a,i}$ because of the electromagnetic interaction 
with the electron background which neutralizes the proton charge. 
In the Wigner-Seitz approximation the functional results \cite{LS91}
\begin{equation}
e_{a,i}= e^0_{a,i}-c_C z^2 a^{-1/3} \left ( \frac{3}{2}\left 
(\frac{\rho_p}{\rho_{0p}}\right )^{1/3}-\frac{1}{2}\left
(\frac{\rho_p}{\rho_{0p}}\right )^{1/2} 
\right ) 
\label{WS}
\end{equation}
where $\rho_p=\rho_e$ is the proton density and $\rho_{0p}\geq \rho_p$ the corresponding saturation value.
The minimal scale at which charge neutrality is verified, is called a Wigner-Seitz cell.

We use in the following a simple liquid-drop parameterization for the cluster energies $e^0_{a,i}$ \cite{starmatter_mmm}:
\begin{equation}
e^0_{a,i}=\left ( -c_V a +c_S a^{2/3}\right ) \left (1-c_I\frac{i^2}{a^2}\right ) +c_C z^2 a^{-1/3}. 
\label{eq:B}
\end{equation}
but shell and pairing corrections can be readily incorporated \cite{hempel}. 
Density dependent correction terms accounting for the nuclear interaction 
with the free nucleons \cite{samaddar,schwenk,heckel,typel} are also expected
to improve the predictive power of the model, as well as a more sophisticated 
form for the cluster internal entropy using realistic $a$ and $i$ dependent
densities of states \cite{starmatter_mmm}. Since none of these improvements is
expected to change the qualitative results of this paper, we are not including 
them here to keep an analytically solvable model.

As a first approximation, one can consider that the the system of interacting nucleons is equivalent to a system of non-interacting clusters, nuclear interaction being completely exhausted by clusterization\cite{fisher}.
This classical model of clusterized nuclear matter is known in the literature as nuclear statistical equilibrium (NSE) \cite{NSE,mishustin,blinnikov,souza}.
This simple model can only describe diluted matter at $\rho \ll \rho_0$ as it can be found in the outer crust of neutron stars,
while nuclear interaction among nucleons and clusters has to be included for applications at higher density, when the average inter-particle distance becomes comparable to the range of the force.

In our model, interactions among composite clusters are taken into account in the simplified form of a hard sphere excluded volume. Since the nuclear density is (approximately) constant inside the clusters, the volume occupied
by each species $(a,i)$ is given by $V_{a,i}=a n_{a,i}/\rho_0$, where
$n_{a,i}$ is the abundance of the species $(a,i)$. The volume fraction available to the clusters then reads: 
\begin{equation}
\frac{V_F}{V}=1-\sum_{a>1,i} a \frac{n_{a,i}}{\rho_0 V}= 
1-\frac{\rho_{cl}}{\rho_0},
\label{freeV}
\end{equation}
where $\rho_{cl}$ is the total density of nucleons bound in clusters.
A slightly different expression has been used in
Refs. \cite{hempel,starmatter_mmm}, 
which however does not change the results presented in this paper.

In addition to the excluded volume effect for the clusters,
the inter-particle nuclear interaction is also considered for the nucleons not bound in clusters.
 The free nucleons self-energy is computed in the 
self-consistent Hartree-Fock approximation with a phenomenological realistic effective interaction \cite{SKM*}. The energy density can be expressed as a function of the density of neutrons $\rho_n$ and protons $\rho_p$ which are not bound in clusters as 
:  
\begin{eqnarray}
\epsilon^{(HM)}&=& \frac{\hbar ^{2}}{2m_0}(\tau _{n}+\tau _{p}) \nonumber \\
&+&t_{0}(x_{0}+2)(\rho _{n}+\rho _{p})^{2}/4
-t_{0}(2x_{0}+1)(\rho _{n}^{2}+\rho_{p}^{2})/4 \nonumber \\
&+& t_{3}(x_{3}+2)(\rho _{n}+\rho_{p})^{\sigma +2}/24
-t_{3}(2x_{3}+1)(\rho _{n}+\rho_{p})^{\sigma }(\rho _{n}^{2}+\rho _{p}^{2})/24 \nonumber \\
&+& \left(t_{1}(x_{1}+2)+t_{2}(x_{2}+2)\right)(\rho _{n}+\rho _{p})(\tau _{n}+\tau_{p})/8
+\left(t_{2}(2x_{2}+1)-t_{1}(2x_{1}+1)\right) (\rho _{n}\tau _{n}+\ \rho _{p}\tau_{p})/8 , \label{skyrme}
\end{eqnarray}

where $m_0$ is the nucleon mass, $t_0,t_1,t_2,t_3, x_0,x_1,x_2,x_3,\sigma$ are Skyrme parameters and $\tau_n,\tau_p$ 
represent the neutron and proton kinetic energy density. 
We use the SKM* \cite{SKM*} parameterization for the numerical applications.

At density $\rho \geq \rho_0$ the whole baryonic matter is expected
to be homogeneous and described by eq.(\ref{skyrme}). The transition from inhomogeneous to homogeneous matter is physically realized in stellar matter at the interface between the crust and the core of a neutron star. As we can see from eq.(\ref{WS}), in our model at proton densities $\rho_p = {\rho_{0p}}$ the Wigner-Seitz correction 
exactly compensates the Coulomb self-energy of the cluster and 
the total Coulomb energy vanishes, reflecting matter homogeneity at 
supersaturation densities.  In this regime the asymptotic cluster energy represents the energy density of homogeneous neutral nuclear matter
%
\begin{equation}
\lim_{a\to\infty}\frac{1}{a}e_{a,i}(\rho_p= \rho_{0p})=\frac{1}{\rho}\epsilon^{HM}(\rho_n,\rho_{0p}),
\end{equation}
where $\rho=\lim_{A,V\to\infty}A/V=\rho_n+\rho_p$, $\rho_I=\lim_{I,V\to\infty}I/V=\rho_n-\rho_p$, 
and $A,I$ are the total number of particles and chemical asymmetry.
The crust-core transition can thus be seen equivalently as the
melting of clusters inside dense homogeneous matter, or as the emergence of a percolating cluster of infinite size.

Charge neutrality is imposed globally but local charge dishomogeneities at the scale of the Wigner-Seitz cell naturally appear in the thermodynamic conditions where matter is clusterized. 
In turn, this gives rise to a long range monopole component of the Coulomb potential 
which extends over domains of the order of the cluster size, and which can 
potentially become macroscopic in the limit of very extended clusters close to
the crust-core transition. As we will show in detail, these long range Coulomb 
correlations are at the origin of the specific thermodynamics.

\subsection{Grandcanonical formulation}
 
Considering that the center of mass of composite fragments can be treated as a 
classical degree of freedom,  their grandcanonical partition sum reads
\cite{starmatter_mmm,hempel}
\begin{equation}
{\cal Z}^{a>1}_{\beta,\mu,\mu_I}=\prod_{a>1,i\epsilon(-a,a)}{\exp{z}_{a,i}} \label{start}
\end{equation}
The partition sum associated to a cluster composed of $n$ neutrons and $z$
protons in a volume $V$ is given by
\begin{equation}
{z}_{a,i}=  V_F  \left ( \frac{2\pi a m_0}{\beta h^2}\right )^{3/2}     
\exp-\left [ \beta \left ( f^\beta_{a,i}-\mu_{a,i} \right )\right ]  \label{z_NZ}
\end{equation}
Here, $V_F$ is the free volume associated to the cluster center of mass given by eq.(\ref{freeV}),
and the cluster chemical potential is a linear
combination of the isoscalar and isovector chemical potentials $\mu$, $\mu_I$ 
which have to be introduced in the presence of two conserved charges
\begin{equation}
\mu_{a,i}=\mu a+\mu_I i.
\end{equation}
Fermi statistics cannot  be neglected when treating $a=1$ fragments 
(protons and neutrons). This component of the baryonic partition sum is thus 
included in the finite temperature Hartree-Fock approximation \cite{starmatter_mmm}
\begin{equation}
  {\cal Z}_{\beta,\mu,\mu_I}^{a=1} = {\cal Z}_{\beta,\mu,\mu_I}^{0} \exp \left [ - \beta \left( \frac{\partial}{\partial{\beta}} \ln {\cal Z}_{\beta,\mu,\mu_I}^{0}  + V\epsilon^{HM} \right) \right ]
  \label{HF}
\end{equation} 
where the non-interacting part of the partition sum can be expressed 
as a functional of the  kinetic energy density $\tau _{q}$ 
for neutrons ($q=n$) or protons ($q=p$) : 

\begin{equation}
\ln {\cal Z}_{\beta,\mu,\mu_I}^{0}  
=\frac{2 \beta V}{3} \sum_{q=n,p} \tau_{q} \frac{\partial \epsilon^{HM}}{\partial \tau_{q}}   
\end{equation}

with:

\begin{equation}
 \tau _{q} = \frac{8\pi}{h^{3}} 
\int_{0}^{\infty } \frac{1}{\hbar^2} 
 \frac{p^{4}dp}{1+e^{\beta ( e_{q}-\mu _{q} )}}
   . \label{EQ:tau}
\end{equation}

The number density of free protons and neutrons 
$\rho_{fq}=\partial_{\beta\mu_q}\ln {\cal Z}^{a=1}_{\beta,\mu,\mu_I}/V$
determines the single particle energy which enters in the self-consistency
equation (\ref{EQ:tau}) according to,

\begin{equation}
e_{q}= p^{2}\frac{\partial \epsilon^{HM}}{\partial \tau_{fq}}   +
\frac{\partial \epsilon^{HM}}{\rho _{fq}}.
\end{equation}

The computation of all thermodynamical variables is straightforward using the 
standard grandcanonical expressions for the global baryonic partition sum 
${\cal Z}={\cal Z}^{a>1}{\cal Z}^{a=1}$. In particular the total baryonic 
pressure is simply $\beta p=\ln {\cal Z}_{\beta,\mu,\mu_I}/V$, 
the multiplicity of the different clusters is given by
\begin{equation}
 n_{a,i} = \frac{\partial \ln {\cal Z}_{\beta,\mu,\mu_I}} {\partial
   \beta\mu_{a,i}} 
= {z}_{a,i} \label{mult}
\end{equation}
and the total baryonic density is the sum of the clusterized ($a>1$) 
and the unbound ($a=1$) component
\begin{equation}
 \rho = \rho_p + \rho_n = \sum_{a,i} a \frac{n_{a,i}}{V} 
 + \rho_{f} =\rho_{cl}
 +\rho_f   \label{dens_tot}
\end{equation}
where $\rho_f=\rho_{fp}+\rho_{fn}$ is the total density associated to the
nucleons which are not bound in clusters.

Expression Eq. (\ref{WS}) includes the electrons self-energy
inside the Wigner-Seitz cell. This means that in the global stellar partition
sum the remaining electron contribution will be factorized 
${\cal Z}_{tot}={\cal Z}^{bar}_{\beta,\mu,\mu_I}{\cal Z}^{el}_{\beta,\mu_e}$,
where ${\cal Z}^{bar}_{\beta,\mu,\mu_I}$ is given by Eq. (\ref{start}) 
and ${\cal Z}^{el}_{\beta,\mu_e}$ is a trivial ideal Fermi gas contribution
which has no influence on the thermodynamics and will not be discussed further
\cite{LS91}.

 
%
\begin{figure}
\begin{center}
\includegraphics[angle=0, width=0.8\columnwidth]{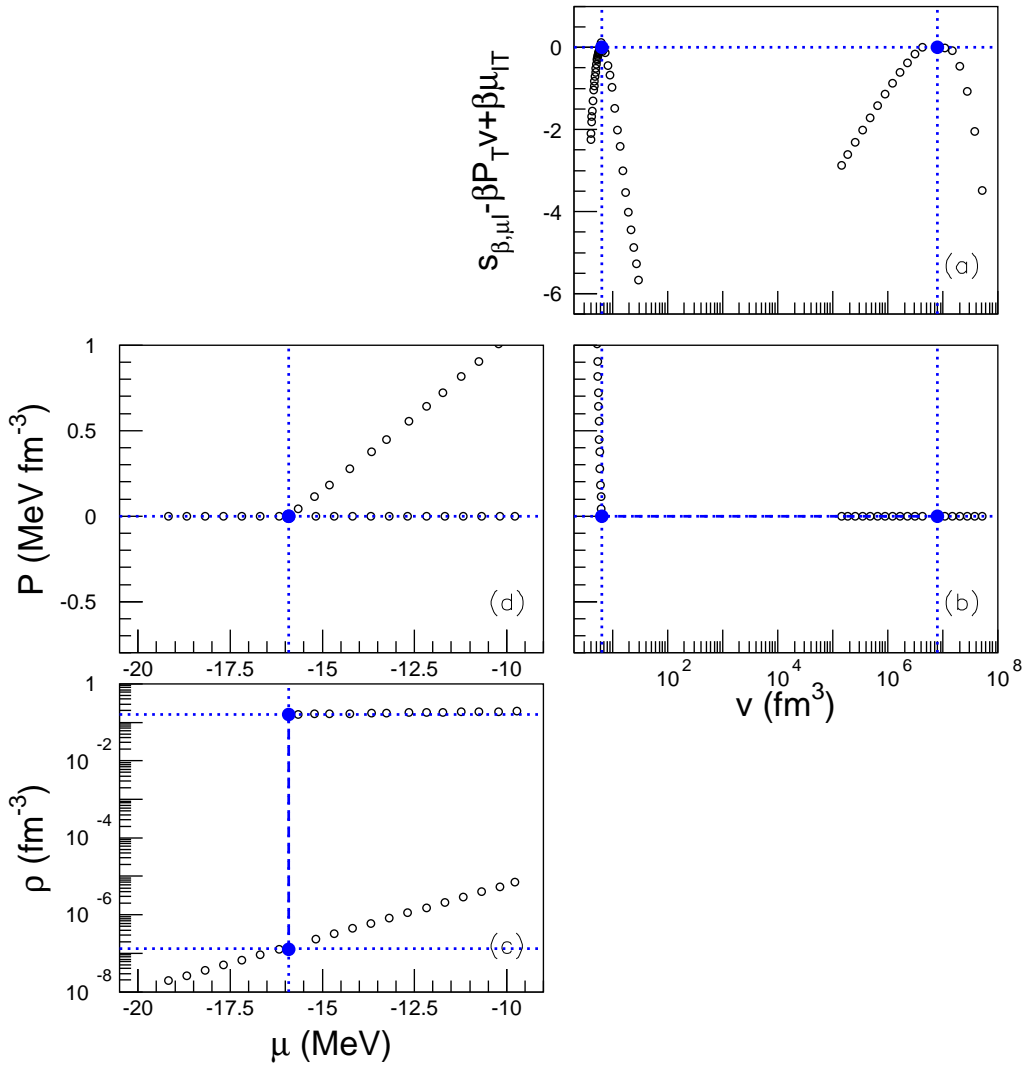}
\end{center}
\caption{Grandcanonical thermodynamics at
$T$=1.6 MeV and $\mu_I$=1.68 MeV. 
Constrained entropy per baryon (a) and associated pressure (b) 
as a function of the inverse baryonic density; to better evidence the concave
behavior of $s_{\beta,\mu_I}$
a linear function $(\beta P_T/\rho-\beta \mu_{IT})$ is subtracted; 
(c) chemical potential as a function of the baryonic density; 
(d) pressure as a function of the chemical potential.
Dashed lines: Gibbs construction. 
}  
\label{fig:thermo_gc}
\end{figure}

The functional relation between the different grandcanonical variables 
is represented in Fig. \ref{fig:thermo_gc} for a representative 
given value of $T$ and $\mu_I$, relevant for the astrophysical applications. 
The constrained entropy  per baryon
$s_{\beta,\mu_I}=\sigma_{\beta,\mu_I}/\rho$ 
is evaluated from the Legendre transform of the partition sum in the region 
where the density is defined:
\begin{equation}
\sigma_{\beta,\mu_I}(\rho) = \ln  {\cal Z}_{\beta,\mu,\mu_I}/V -\beta\mu\rho \label{legendre},
\end{equation}
and shown in panel (a) as a function of the volume per baryon or 
inverse density $v=1/\rho$. The first derivative of this function is the
baryonic pressure $P$ given in panel (b). 
The first derivative of the entropy density $\sigma_{\beta,\mu_I}$  
with respect to the density gives the other equation of state,
 namely the chemical potential  $\mu=P/\rho-s_{\beta,\mu_I}/\beta$, 
represented in panel (c). Finally phase equilibrium is best spotted by 
looking at the phase diagram given by the relation between intensive
variables, as shown in panel (d).

The grandcanonical thermodynamics leads to a first order phase transition. 
This can be inferred  from the characteristic two-humped structure of the 
constrained entropy, which is better evidenced by subtraction of a straight
line, as well as by the crossing of the two equations of state (panel (d)).  
In the presence of a first order phase transition only in the ensemble where 
the order parameter is fixed (here: the canonical ensemble) the 
phase transition region is accessible, while it is jumped over in the ensemble 
where the order parameter is fixed only in average (here: the grandcanonical).
As a consequence, a discontinuity is observed in the equations of state
covering a huge range of baryonic densities relevant for the 
description of supernova dynamics.

 The equilibrium solution in the phase transition region corresponds to a 
linear combination of the two pure phases following Gibbs rules.
This construction is exactly equivalent to a one-dimensional Maxwell
construction if we work  in an ensemble where all intensive parameters
are fixed but one, and it is represented by the dashed lines in 
Fig. \ref{fig:thermo_gc} \cite{glendenning,pagliara}.
We note on passing that this simplification demands to work in the 
non-standard ensemble $(T,\rho,\mu_I)$. In astrophysical applications 
it is customary to work rather with the parameters $(T,\rho,y_p)$, 
where $y_p=\rho_p/\rho$ is the proton fraction. Within this ensemble, 
a full two-dimensional Gibbs construction is needed to correctly calculate 
the coexistence zone. The use of a Maxwell construction in the ensemble 
$(T,\rho,y_p)$ is a mistake, yet it is often used in the literature \cite{hempel,LS91}.  

The construction of a convex entropy envelope allows to recognize that inside
the density region, indicated by dotted lines, the obtained grandcanonical
solutions do not correspond to an equilibrium, since a higher entropy solution 
can be obtained by making a linear combination of the two (dense and diluted) 
solutions marked by a filled circle, that is by a first order phase transition. 
The pressure and chemical potential associated to the metastable branch on the 
low density side are also represented in Fig. \ref{fig:thermo_gc}. 
From the pressure viewpoint (panel (b)) such solutions appear equivalent 
to a phase coexistence (dashed line), but this is not true in 
terms of chemical potential (panel (c)).
 
The presence of metastable solutions was never discussed in the framework of
NSE models \cite{NSE,mishustin,blinnikov,souza,hempel} to our knowledge. 
It can be understood from the fact that, for chemical potentials higher 
than the Fermi energy of dense uniform matter $\mu\geq e_F\approx -16 MeV$, 
the equilibrium condition can be obtained either as a mixture of 
clusters and homogeneously distributed nucleons $\mu^{a>1}=\mu^{a=1}=\mu$, 
or alternatively letting the clusterized component to vanish 
($\mu^{a=1}=\mu$ and $\rho_{cl}=0$). This gives a second stationary entropy
solution which appears to correspond to the absolute entropy maximum, 
and which renders metastable the clusterized solution at lower density.
 
The ending point of the metastable branch can be worked out easily.
Indeed from Eq. (\ref{z_NZ}) we can see that, for any temperature $\beta^{-1}$ 
and isovector chemical potential $\mu_I$, it exist a limiting value of 
chemical potential $\mu_{max}$ for which the cluster multiplicities 
$n_{A,I}$ asymptotically diverge. This value is determined by the condition 
of an asymptotically negative Gibbs free energy
\begin{equation}
\lim_{a\to\infty} f^\beta_{a,i}-\mu a - \mu_I i < 0
\end{equation}
combined with the requirement that the Coulomb energy vanishes in the limit 
of an infinitely extended homogeneous cluster:  
\begin{equation}
\frac{1}{V}\sum_{a,i}\frac{a-i}{2}n_{a,i}=\rho_{0p}
\end{equation}
For chemical potentials above $\mu_{max}$, clusterized partitions become
unstable due to the emergence of a liquid phase.

Coming back to the discontinuity shown in the equations of state in Figure \ref{fig:thermo_gc}, 
in standard thermodynamics the fact that the Gibbs construction is a
posteriori made to fill up the coexistence region is not a limitation, 
because ensemble equivalence guarantees that the very same linear combination 
solution would have been obtained if we had worked in the canonical ensemble, 
explicitly constraining the density.  

Because of the complexity of the phenomenology of stellar matter we have not
proposed an Hamiltonian treatment of the problem. 
However the phenomenological free energy functional Eq. (\ref{eq:B})
we have used implicitly contains the effect of the attractive short range
nuclear forces, scaling proportionally to the number of particles in the 
thermodynamic limit, and repulsive long range Coulombic forces, 
scaling proportionally to the square of the number of particles and 
only partially screened. None of the possible improvements on the nuclear
energy functional would change these very general scaling behaviors. 
It has been argued in ref.\cite{ising_star}
that this generic frustration effect should lead to a quenching of the 
first order phase transition. We have shown that the phase transition is 
observed in the grandcanonical ensemble. We turn therefore to explore the 
possibility that ensemble in-equivalence might be at play in the stellar 
environement, with the thermodynamic anomalies associated to the
thermodynamics of long range interactions \cite{dauxois}. 
  
\subsection{Canonical formulation}
  
A fully canonical formulation of our model would imply the use of two
independent extensive variables, the proton and neutron density 
$(\rho_p,\rho_n)$ or equivalently the isoscalar and isovector density
$(\rho,\rho_I)$. 
It is however well established \cite{chomaz,ducoin} that, contrary to other 
physical systems like binary alloys and molecular mixtures where  phase
transitions can also imply separation of the species \cite{sivardiere},
the isovector density is not an order parameter of the nuclear matter phase transition.
Since the in-equivalence effect we are looking for is associated to the
phenomenon of phase coexistence, we thus expect to see it even if we 
keep a grandcanonical treatment for the isovector density. 

The effect we are looking for is due to the long range Coulomb interaction, 
which vanishes in homogeneous matter. This Coulomb interaction is neglected 
in the computation of the abundances of $a=1$ particles which are modelized 
as homogeneously distributed. For this reason we will also stick to a
grandcanonical formulation for $a=1$ particles according to Eq. (\ref{HF}), 
and assume that the approximate Legendre transformation Eq. (\ref{legendre})
\begin{equation}
 {\ln {\cal Z}_{\beta,\mu_I}^{a=1}(\rho)} = {\ln {\cal Z}_{\beta,\mu,\mu_I}^{a=1} } -\beta\mu\rho V,
\end{equation}
which is exact in the mean-field approximation we have employed, is physically correct.
%
%
%
%
For the same reason we will consider that the standard thermodynamic
assumption of total canonical entropy being additive among independent
components, is verified as expected at the thermodynamic limit within ensemble equivalence: 
\begin{equation}
\sigma_{\beta,\mu_I}^{can} (\rho)= {\ln {\cal Z}_{\beta,\mu_I}^{a=1}(\rho_f)}   + \lim_{V\to\infty} \frac{1}{V}\ln {\cal Z}_{\beta,\mu_I}^{a>1}\left(V \rho_{cl} \right) \label{monomer_mix}
\end{equation}
where the density repartition between the clustered $\rho_{cl}$ 
and unbound $\rho_f$ component Eq. (\ref{dens_tot}) is uniquely defined by the
condition of having a single chemical potential $\mu$ for both components.
Possible extra deviations from ensemble equivalence originating from the $a=1$ 
contribution would need a more sophisticated model where the polarization 
of the free protons would be explicitly accounted for \cite{hempel}.
For simplicity, in the following we will refer to this hybrid
$(\beta,\mu_I,\rho)$ 
ensemble as to the "canonical" ensemble.

To derive the expression of the canonical partition sum, we start from the 
general statistical mechanics relation which links the different statistical
ensembles, restricted to composite clusters $a>1$ only
(the subscript $a>1$ is omitted hereafter for simplicity):
\begin{equation}
{\cal Z}_{\beta,\mu,\mu_I}=\sum_{A>1} {\cal Z}_{\beta,\mu_I}(A) \exp \beta\mu A
\end{equation}
Identification with Eq. (\ref{start}) gives,
\begin{equation}
{\cal Z}_{\beta,\mu_I}(A)=\sum_{\{n_a\}}\prod_{a>1}{\frac{\omega_a^{n_a}}{n_a!}} \label{cano}
\end{equation}
where $n_a=\sum_i n_{ia}$ is the occupation number of size $a$ and the sum 
is restricted to combinations   $\{n_a\}\equiv \{n_2,\dots,n_A \}$
satisfying the canonical constraint,
\begin{equation}
\sum_{a=2}^A a n_a=A. \label{canoconstr}
\end{equation}
The weight of the different cluster size is given by,
\begin{equation}
\omega_a= \sum_{z=0}^a \omega_{a,a-2z} \exp (\beta \mu_I (a-2z)), \label{isotopes}
\end{equation}
and the weight of each nuclear species is determined by the 
free energy functional we have assumed,
\begin{equation}
\omega_{a,i} =  V_F  \left ( \frac{2\pi a m_0}{\beta h^2}\right )^{3/2}     
\exp-\left ( \beta  f^\beta_{a,i} \right ) .
\end{equation}
Following Ref. \cite{mekjan} we introduce an auxiliary canonical partition sum 
${\cal Z}_{\beta,\mu_I}^m(A)$ defined by the additional constraint that the 
cluster multiplicity is fixed to $m$, $\sum_a n_a=m$. 
With the help of relation (\ref{cano}) we get,
\begin{equation}
{\cal Z}_{\beta,\mu_I}^{m-1}(A-a)=\frac{<n_a>_m}{\omega_a}{\cal Z}_{\beta,\mu_I}^{m}(A),
\end{equation}
where $<n_a>_m$ is the average multiplicity of size $a$ under the additional 
constraint of total multiplicity $m$. Explicitly implementing the canonical 
constraint Eq. (\ref{canoconstr}) we arrive to the recursion relation, 
\begin{equation}
{\cal Z}_{\beta,\mu_I}^{m}(A)= \frac{1}{A}\sum_{a=2}^A a \omega_a {\cal Z}_{\beta,\mu_I}^{m-1}(A-a),
\end{equation}
which is valid for any value of $A$. Summing over all the possible $m$ values, 
we can see that the same
relation holds for the unconstrained canonical partition sum
\begin{equation}
{\cal Z}_{\beta,\mu_I}(A)= \frac{1}{A}\sum_{a=2}^A a \omega_a {\cal
  Z}_{\beta,\mu_I}(A-a) .
\label{zcan}
\end{equation}
This expression can be recursively computed with the initial condition 
${\cal Z}_{\beta,\mu_I}(2)=\omega_2$.
The computational problem however arises that going towards the thermodynamic limit
the evaluation of the double sum implied by Eq. (\ref{isotopes}) becomes numerically very heavy.
Starting from a sufficiently high minimal value $a_{min}$ 
we therefore develop a continuous approximation for Eq. (\ref{isotopes}),
\begin{equation}
\omega_{a>a_{min}} \approx \frac{1}{2}\int_{-a}^a dy \exp g(y),
\end{equation}
This integral is calculated in the saddle point approximation:
\begin{equation}
g(i)=\ln \omega_{a,i} +\beta \mu_I i \approx \ln \omega_{a,<i>}+
\beta\mu_I <i> - \frac{(i-<i>)^2}{2\sigma^2_a}.
\end{equation}
where the most probable isotopic composition $<i>$ of a cluster of size $a$ 
depends on the temperature according to,
\begin{equation}
\mu_I=\frac{\partial f_{a,i}}{\partial i}|_{i=<i>},
\end{equation}
and the associated dispersion is given by
\begin{equation}
\frac{1}{\sigma^2_a}=\beta \frac{\partial^2 f_{a,i}}{\partial i^2}|_{i=<i>}.
\end{equation}
The global weight of size $a$ finally results,
\begin{equation}
\omega_{a>a_{min}} \approx \omega_{a,<i>}\sqrt{2\pi\sigma^2_a} 
\exp (\beta\mu_I<i>)/2 \label{saddle}
\end{equation}
and the value of $a_{min}$ is chosen such that Eq. (\ref{saddle}) and
Eq. (\ref{isotopes})  give 
estimation differing less than 5\%.
The canonical thermodynamical potential $\sigma_{\beta,\mu_I}^{can}$ 
is then defined by Eq. (\ref{monomer_mix}), with the identification 
${\cal Z}_{\beta,\mu_I} \equiv {\cal Z}_{\beta,\mu_I}^{a>1}$.
%
%

Ensembles are  equivalent if this function 
 coincides with the entropy density $\sigma_{\beta,\mu_I}(\rho)$ 
defined by the Legendre transform of the grandcanonical partition sum 
Eq. (\ref{legendre}). More precisely, ensemble equivalence demands that the 
thermodynamic quantities as calculated from the canonical partition sum,
\begin{eqnarray}
\mu_{can} &=& -\frac{1}{\beta}\frac{\partial \sigma_{\beta,\mu_I}^{can}  }{\partial \rho}, \nonumber \\
p_{can} &=& \frac{\rho \sigma_{\beta,\mu_I}^{can}}{\beta}+\mu\rho, \nonumber \\
\epsilon_{can} &=& -\frac{\partial  \sigma_{\beta,\mu_I}^{can}}{\partial \beta}, 
\end{eqnarray}
coincide with the corresponding grandcanonical quantities.
\begin{figure}
\begin{center}
\includegraphics[angle=0, width=0.9\columnwidth]{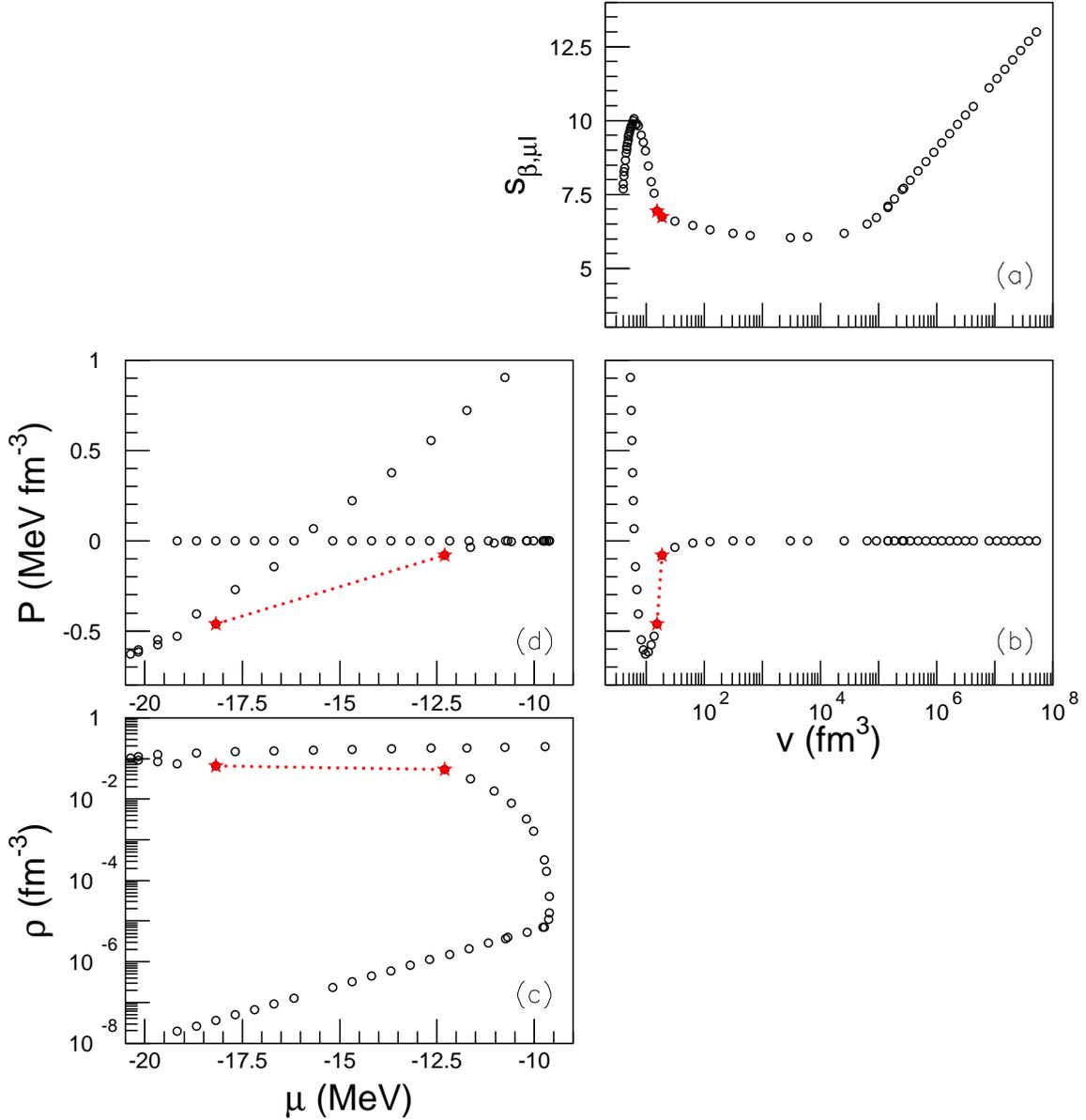}
\end{center}
\caption{Constrained entropy (a), pressure (b),(c) and chemical potential
  (b)(d) evaluated in the canonical ensemble in the same thermodynamic
  conditions as in Fig. 1. 
The pressure and chemical potential discontinuity is indicated by a dotted line.  
}
\label{fig:anomaly}
\end{figure}
The canonical thermodynamics, in the same thermodynamic conditions as in 
Fig. \ref{fig:thermo_gc}, is displayed in Fig. \ref{fig:anomaly}.
We can see that the canonical calculation allows to interpolate between the
dense and diluted branches observed in the grandcanonical ensemble as
expected. However the interpolation is not linear, meaning that
the chemical potential continuously varies as a function of the density. 
The discontinuity in the entropy slope at high density leads to a jump in the 
intensive observables close to the saturation density, in complete
disagreement with the grandcanonical solution \cite{dauxois}.
Even more interesting, the entropy presents a convex intruder, the behavior of
the equations of state is not monotonous and a clear back-bending is observed, 
qualitatively similar to the phenomenon observed in first order phase
transitions in finite systems \cite{gross}.

\begin{figure}
\begin{center}
\includegraphics[angle=0, width=0.45\columnwidth]{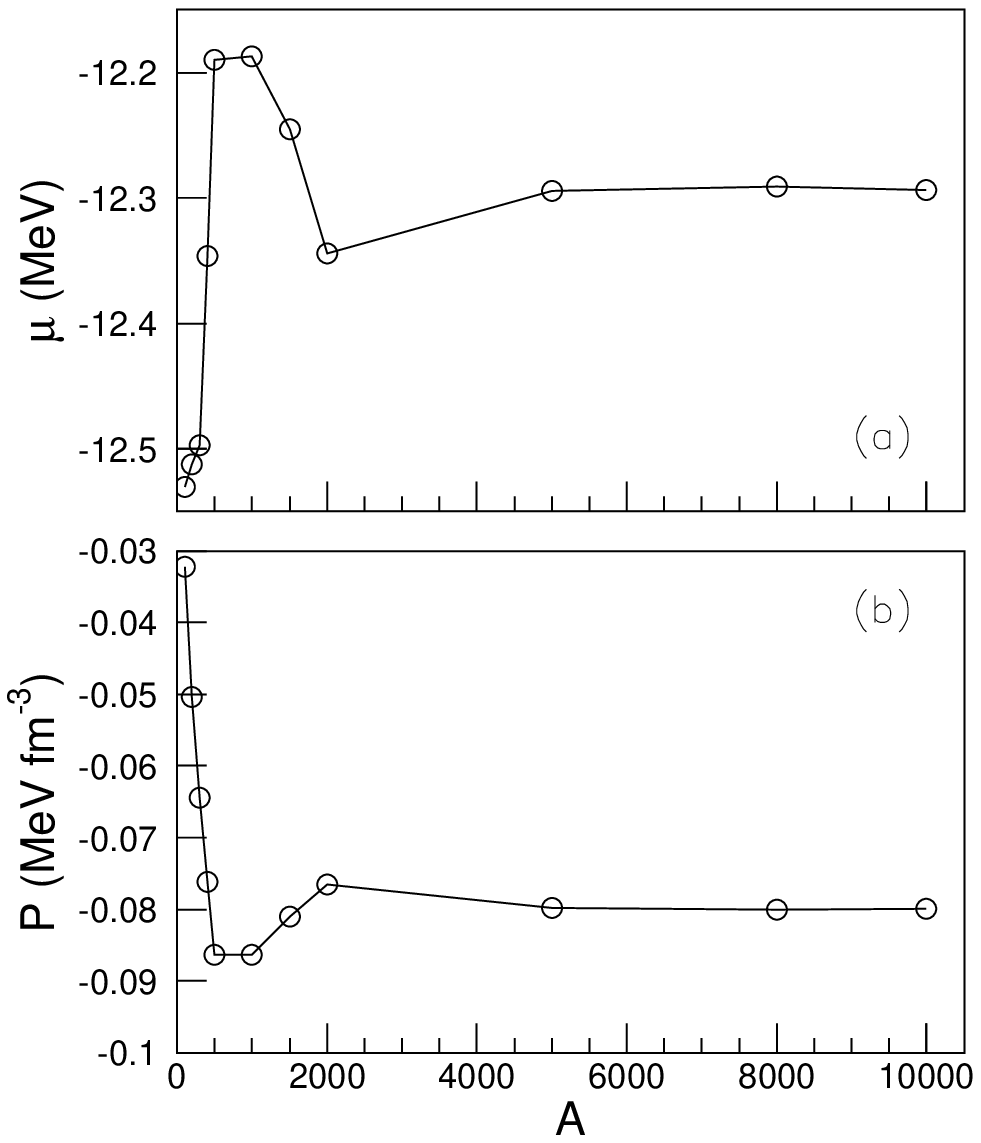}
\includegraphics[angle=0, width=0.45\columnwidth]{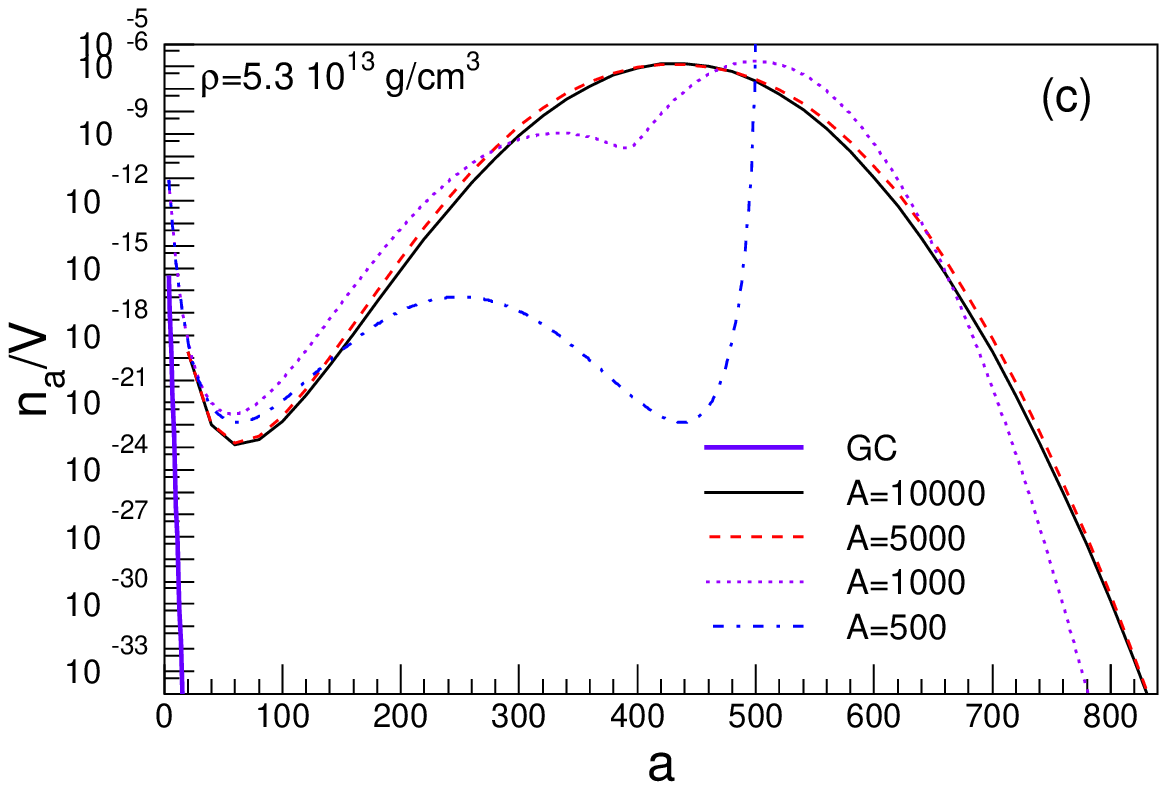}
\end{center}
\caption{Convergence study of the canonical ensemble. The baryonic chemical
  potential (a), 
baryonic pressure (b) and cluster multiplicity per unit volume  as a function
of the cluster size are represented at a representative density inside the
ensemble in-equivalence region,  by varying the total system size.  
Thick line in panel (c):  cluster multiplicity distribution in the 
grandcanonical ensemble.  
}
\label{fig:convergence}
\end{figure}
The canonical partition sum Eq. (\ref{zcan}) is only defined for finite values
of the total number of particles $A$.
The doubt therefore arises  that this non-trivial behavior and the similarity 
with the finite systems thermodynamics  might be due to a non achievement 
of the thermodynamic limit.
Fig. \ref{fig:convergence} shows that this is not the case, and the canonical
calculation is convergent. In this figure the chemical potential (a), 
pressure (b) and cluster size distribution (c) are represented 
for different values of the total number of particles $A$ 
at a given density inside the in-equivalence region. 
  
We can see that indeed a very large size has to be used before this limit is
achieved. As a rule of thumb, the total system size has to typically be
approximately ten times bigger than the most probable cluster size
in order to have convergent results, meaning that the equivalent of a
Wigner-Seitz cell contains in average $\approx 10$ dominant clusters.  
This can be understood considering that at finite temperature the distribution 
is very large, and the often used single-nucleus approximation \cite{LS91} is not realistic. 

As it can be seen in Fig. \ref{fig:convergence}, the grandcanonical
equilibrium prediction for this thermodynamic condition would correspond to a 
macroscopic liquid fraction in equilibrium with essentially free particles 
(dashed line). An explicit computation of the coexistence region in the
canonical ensemble shows that this is not the case, as the cluster sizes do
not scale with the total system size $A$. As one may notice, this liquid
fraction is replaced by a finite, though large, nucleus with a characteristic 
radius of the order of only $5-10$ fm.

This finding is in agreement with all the theoretical microscopic modelizations
which are naturally elaborated inside a single Wigner-Seitz cell within a
fixed number of particles \cite{horowitz,watanabe_prl,newton,sebille}. 
All these canonical studies agree in predicting that matter is clusterized 
for all subsaturation densities and the cluster size and composition 
evolve continuously with the density, which is incompatible with a
modelization based on phase coexistence \cite{ohnishi} where only the relative 
proportion of the two phases varies through the phase transition.

As it has been argued in Refs. \cite{ising_star}, the quenching of the phase 
transition is due to the high electron incompressibility. Because of the 
charge neutrality constraint over macroscopic distances, an intermediate 
density solution given by a linear combination of a high density homogeneous 
region and a low density clusterized region would imply an infinite repulsive 
interaction energy due to the electron density discontinuity at the 
(macroscopic) interface \cite{ising_star}. 

This argument requires electrons to be completely incompressible. 
Since the electron incompressibility, while high, is not infinite, 
it could have been argued that a slightly modified coexistence region should
be observed, where the liquid fraction would be constituted by large but still 
mesoscopic clusters, such that the interface energy would not diverge. 
The comparison between canonical and grandcanonical shown in
Fig. \ref{fig:thermo_gc}, \ref{fig:anomaly}, \ref{fig:convergence} 
demonstrates that this is not true: the presence of microscopic, 
instead than macroscopic, fluctuations, qualitatively change both the
thermodynamics and the composition of matter.

\begin{figure}
\begin{center}
\includegraphics[angle=0, width=0.4\columnwidth]{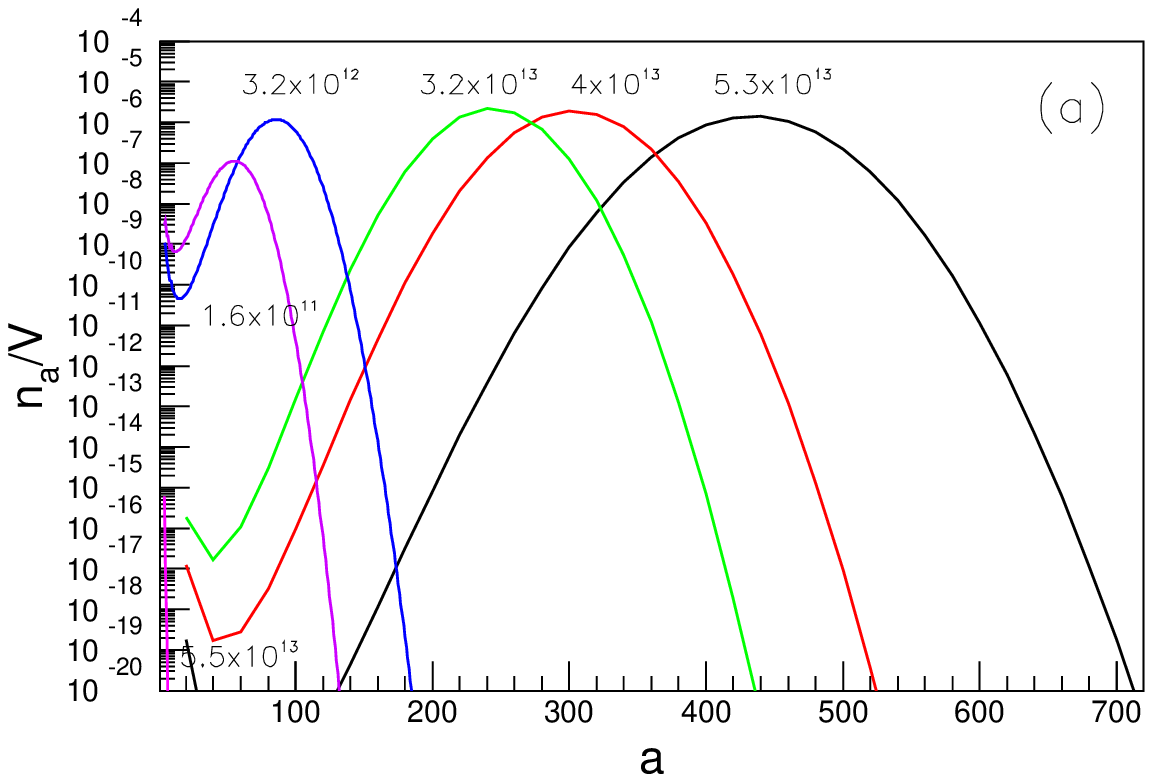}
\includegraphics[angle=0, width=0.4\columnwidth]{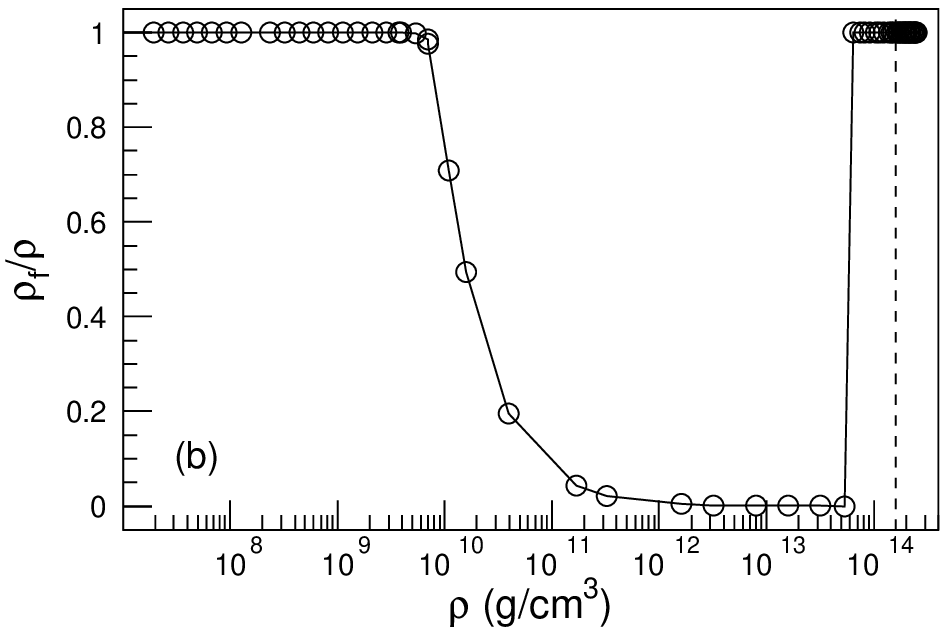}
\end{center}
\caption{ (a) Cluster distributions 
as a function of the density in the ensemble in-equivalence region;
(b) mass fraction of unclusterized matter as a function of the total
baryonic density, in the same thermodynamic conditions as in (a).  
The dashed line gives the saturation density of nuclear matter.
}
\label{fig:transition}
\end{figure}

\section{III. Phenomenological consequences of ensemble in-equivalence}

It is important to remark that in the density region where the 
grancanonical ensemble is defined, the predictions of the ensembles coincide. The in-equivalence is observed in the intermediate density
region, where the grandcanonical first order phase transition is not observed in the canonical ensemble. Since in this density domain
the grandcanonical ensemble is not defined, there is no ambiguity on which ensemble should be chosen, meaning that the phenomenology of star matter has to be described with canonical thermodynamics.

A closer look at the cluster distribution in the in-equivalence region can be
obtained from the left part of Fig. \ref{fig:transition}, which displays 
the cluster distributions 
as a function of the density in the ensemble in-equivalence region.
We can see that this distribution varies continuously, as it is expected
physically, with very heavy nuclei and a very wide size distribution at the
highest densities corresponding to the inner crust. 
These very massive nuclei disappear at a density close to saturation density,
 which defines the density corresponding to the crust-core transition in the model.
This sudden process is at the origin of the non differentiable point observed
in the entropy in Fig. \ref{fig:anomaly}. The resulting discontinuity in
chemical potential and pressure is therefore 
a physical effect in the framework of this model. A word of caution is however necessary.
It is well known from star matter literature \cite{Lattimer85} that close to
saturation density deformed extended
nuclei are energetically favored, the so called "pasta" phases.
We expect that adding a deformation degree of freedom to the cluster energy 
functional would smooth this discontinuity.  

The right part of Fig. \ref{fig:transition} represents the mass fraction of 
unclusterized matter (free nucleons and homogeneous dense matter) as a 
function of the density. Homogeneous matter dominates at the very low
densities which physically correspond to the neutron star atmosphere, 
while clusters become increasingly dominant at higher density, until they melt 
into the homogeneous liquid core. Even in the absence of a first order phase
transition, a very sharp behavior is obtained  defining a relatively precise
value for the crust-core transition density .
 
\begin{figure}
\begin{center}
\includegraphics[angle=0, width=0.75\columnwidth]{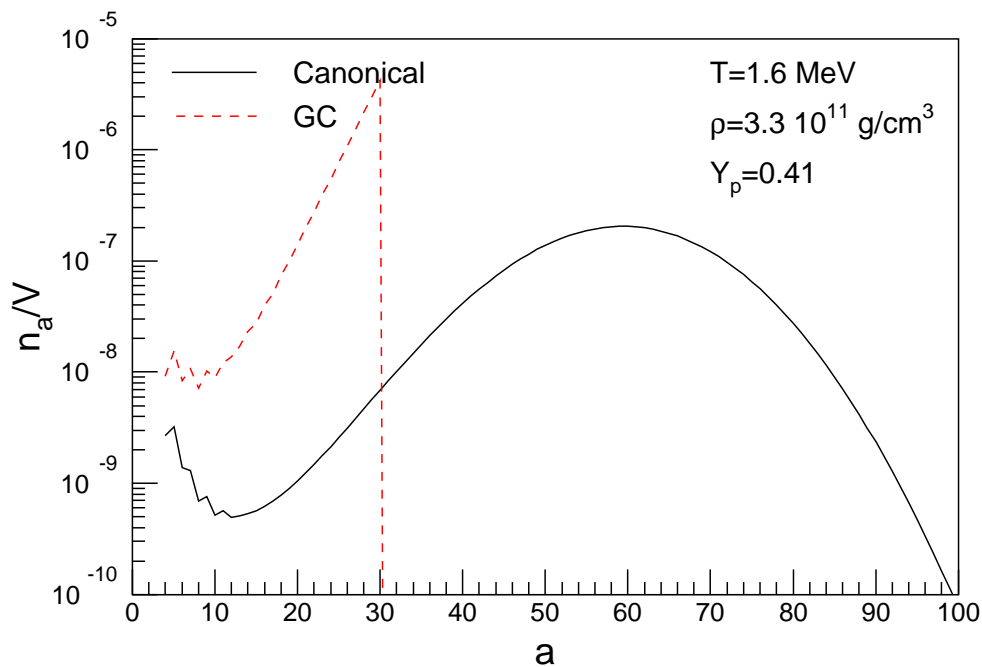}
\end{center}
\caption{Comparison between canonical (full line) and grandcanonical (dashed
  line) predictions for the cluster distribution in a specific thermodynamic 
condition relevant for supernova dynamics. 
}
\label{fig:ecapture}
\end{figure}

Besides the relevance of the issue of ensemble in-equivalence from the 
statistical physics viewpoint, it is interesting to remark that the use of 
grandcanonical thermodynamics can lead to important qualitative and
quantitative discrepancies in the computation of different physical quantities
of interest for the astrophysical applications.  This is demonstrated in 
Fig. \ref{fig:ecapture}, which shows the cluster distribution for a chosen 
thermodynamic condition ( temperature $T=1.6$ MeV, baryonic density 
$\rho=3.3\cdot 10^{11}$ gcm$^{-3}$, proton fraction $Y_p=0.41$) which is typical 
for the dynamics of supernova matter after the bounce and before the 
propagation of the shock wave \cite{kitaura,marek}. We can see that the
dominant cluster size is around $A=60$, which is a particularly important 
size in the process of electron capture which determines the composition of
the resulting neutron star
\cite{bethe79,zeldovich,horowitz_nu,pinedo2006,janka,sonoda}. 
It is clear that it is very important to correctly compute the abundances of such nuclei. 

Conversely in a grandcanonical formulation, as the widely used nuclear
statistical equilibrium (NSE) \cite{NSE,mishustin,blinnikov,souza},
these partitions are simply not accessible as they fall in the phase transition region.
An approach consisting in taking the metastable grandcanonical prediction and 
considering only the nuclei of size such that the chosen total density is
obtained, 
as in Ref. \cite{hempel}, is shown by the dashed line in
Fig. \ref{fig:ecapture}. 
It is clear that such approach completely misses the correct cluster distribution.
Alternatively, hybrid canonical-grandcanonical formulations are routinely used
in the astrophysical community \cite{LS91,shen,shen_horowitz}. 
Such approaches do not share the drawback of grandcanonical NSE calculations, 
but they always introduce artificial Maxwell constructions to fill the high 
density part of the equation of state. Moreover, they never address the 
fluctuations in the cluster composition, clusterized matter being modelized 
by a single representative nucleus. It is clear from Fig. \ref{fig:ecapture} 
that this approximation is highly questionable at finite temperature, 
where distributions are wide, the largest cluster does not coincide with the
average one nor with the most probable.   
The model we have presented overcomes all these problems.
It is clear that many improvements are still necessary in this model before it 
can be considered as a reliable quantitative prediction for astrophysical
simulations. 
Both improvements on the cluster energy functional and inclusion of
deformation degrees of freedom are in progress to this aim. However we believe 
that the main results presented in this paper, namely the absence of phase
transition due to Coulomb frustration and the dominance of microscopic
clusters with a large and continuous distribution of size extending over most 
of the subsaturation region, are general results which will not change with a 
more sophisticated model. 

\section{IV. Conclusions}

To conclude, in this paper we have shown that dense stellar matter as it can
be found in core-collapse supernova and in the crust of neutron stars is a 
macroscopic physical example of ensemble in-equivalence.
A first-order phase transition is observed in the grandcanonical ensemble, but when the region 
corresponding to the discontinuity is explored explicitly constraining the 
density in the canonical ensemble, the macroscopic dishomogeneities associated
to phase coexistence are seen to be replaced by microscopic dishomogeneities 
leading to cluster formation. As a consequence, the transition 
observed in the physical system is continuous.
This phenomenon is due to the long range Coulombic interactions which quench 
 the phase transition, thus giving rise to a thermodynamics qualitatively
 similar to the one of finite systems including thermodynamic anomalies. 
Specifically, the relation between density and chemical potential is non
monotonous implying negative susceptibility.

Accounting for this specificity can have sizeable effects in the computation
of different quantities of interest for supernova dynamics.

\begin{acknowledgments}
{This paper has been partly supported by ANR under the project NEXEN 
and by IFIN-IN2P3 agreement nr. 07-44.
Ad. R. R acknowledges partial support from the Romanian National Authority for Scientific
Research under grant PN-II-ID-PCE-2011-3-0092 and kind
hospitality from LPC-Caen.}
\end{acknowledgments}

\end{document}